\documentclass[aoas,MSNbibl,nameyear,seceqn,dvips]{arximspdf}
\usepackage{dcolumn,multirow}
\usepackage{graphicx}

\doi{10.1214/13-AOAS661} 
\volume{7}
\issue{4}
\pubyear{2013}
\firstpage{2138}
\lastpage{2156}

\makeatletter

\newcolumntype{d}[1]{D{.}{.}{#1}}
\def\cal{\mathcal}

\newtheorem{theorem}{Theorem}

\newcommand{\by}{\mathbf{y}}

\newcommand{\bM}{\mathbf{M}}
\newcommand{\cD}{{\cal D}_\mathrm{obs}}
\newcommand{\blambda}{\bolds{\lambda}}
\makeatother

\begin{document}
\begin{frontmatter}

\title{Bayesian data augmentation dose finding with~continual
reassessment method and delayed~toxicity}
\runtitle{Bayesian data augmentation dose finding}

\begin{aug}
\author[a]{\fnms{Suyu} \snm{Liu}\thanksref{t1}\ead[label=e1]{syliu@mdanderson.org}},
\author[b]{\fnms{Guosheng} \snm{Yin}\thanksref{t2,t3}\ead[label=e2]{gyin@hku.hk}}
\and
\author[a]{\fnms{Ying} \snm{Yuan}\corref{}\thanksref{t1,t4}\ead[label=e3]{yyuan@mdanderson.org}}
\affiliation{MD Anderson Cancer Center\thanksmark{t1} and University of
Hong Kong\thanksmark{t2}}
\thankstext{t3}{Supported in part by a Grant (784010) from the Research
Grants Council of Hong Kong.}
\thankstext{t4}{Supported in part by the National Cancer Institute
Grant R01CA154591-01A1.}
\runauthor{S. Liu, G. Yin and Y. Yuan}
\address[a]{S. Liu\\
Y. Yuan\\
Department of Biostatistics\\
University of Texas\\
\quad MD Anderson Cancer Center\\
Houston, Texas 77030\\
USA \\
\printead{e1}\\
\phantom{E-mail:\ }\printead*{e3}} 

\address[b]{G. Yin\\
Department of Statistics\\
\quad and Actuarial Science\\
University of Hong Kong\\
Hong Kong\\
China \\
\printead{e2}} 
%
\end{aug}

\received{\smonth{2} \syear{2013}}
\revised{\smonth{5} \syear{2013}}

%
\begin{abstract}
A major practical impediment when implementing adaptive dose-finding
designs is that
the toxicity outcome used by the decision rules may not be observed
shortly after the initiation of the treatment. To address this issue,
we propose
the data augmentation continual reassessment method (DA-CRM) for dose finding.
By naturally treating the unobserved toxicities as missing data, we show
that such missing data are nonignorable in the sense that the missingness
depends on the unobserved outcomes.
The Bayesian data augmentation approach is used to
sample both the missing data and model parameters
from their posterior full conditional distributions.
We evaluate the performance of the DA-CRM
through extensive simulation studies and also compare it with
other existing methods. The results show
that the proposed design satisfactorily resolves the
issues related to late-onset toxicities and possesses desirable
operating characteristics:
treating patients more safely and also selecting
the maximum tolerated dose with a higher probability.
The new DA-CRM is illustrated with two phase I cancer clinical trials.
\end{abstract}

%

\end{frontmatter}
%
\section{Introduction}\label{secintro}
The continual reassessment method (CRM) proposed by O'Quigley, Pepe and
Fisher (\citeyear{OQuPepFis90}) is an
influential phase I clinical trial design for finding the maximum
tolerated dose (MTD) of a new drug.
The CRM assumes a single-parameter working dose--toxicity
model and continuously updates the estimates of the toxicity probabilities
of the considered doses to guide dose escalation.
Under some regularity conditions, the MTD identified by the
CRM generally converges to the true MTD, even when the working
model is misspecified [\citet{SheOQu96}].
A variety of extensions of the CRM have been proposed to
improve its practical implementation and operating characteristics
[Goodman, Zahurak and Piantadosi (\citeyear{GooZahPia95}); M\"{o}ller
(\citeyear{Mo95}); Heyd and
Carlin (\citeyear{HC99}); Leung and Wang (\citeyear{LeWa02});
\citet{OQuPao03};
\citet{Gar06}; \citet{IasOQu11}; among others]. Recently,
several robust versions of the CRM have been
proposed by using the Bayesian model averaging and posterior maximization
[\citet{YinYua09} and Daimon, Zohar and O'Quigley (\citeyear
{DaiZohOQu11})], so that
the method is insensitive to the prior specification of the
dose--toxicity model.

In real applications, to achieve its best performance, the CRM requires
that the toxicity outcome be observed quickly
such that, by the time of the next dose assignment,
the toxicity outcomes of the currently treated patients
have been completely observed. However, late-onset toxicities are
common in phase I clinical trials, especially in oncology areas. For
example, in radiotherapy
trials, dose-limiting toxicities (DLTs) often occur
long after the treatment is finished [\citet{CoiMyeTep95}].
\citet{Desetal07} reported a phase I study to determine the MTD of
oxaliplatin for
combination with gemcitabine and the concurrent radiation therapy
in pancreatic cancer. In that trial, on average, a new patient arrived every
two weeks, whereas it took nine weeks to assess the toxicity outcomes
of the patients
after the treatment is initiated.
Consequently,
at the moment of dose assignment for a newly arrived patient,
the patients under treatment
might not have yet completed the full assessment period and,
thus, their toxicity outcomes might not be available for making the
decision of dose assignment.
Late-onset toxicity has been becoming a more
critical issue in the emerging era of the development of novel
molecularly targeted agents,
because many of these agents tend to induce late-onset toxicities.
A recent review paper in the \textit{Journal of Clinical Oncology} found
that among a total of 445 patients involved in 36 trials,
57\% of the grade 3 and 4 toxicities were late-onset and,
as a result, particular attention has been called upon the issue of
late-onset toxicity [\citet{Posetal11}].

Our research is motivated by one of the collaborative projects, which
involves the combination
of chemo- and radiation therapy. The trial aims to
determine the MTD of a chemo-treatment, while the radiation therapy is delivered
as a simultaneous integrated boost in patients with locally
advanced esophageal cancer. The DLT is defined as CTCAE 3.0
(Common Terminology Criteria for Adverse Events version 3.0)
grade 3 or 4 esophagitis, and the
target toxicity rate is 30\%. In this trial, six dose levels are
investigated and toxicity is expected to be late-onset.
The accrual rate is approximately 3
patients per month, but it generally takes 3 months to fully assess
toxicity for each patient. By the time of dose assignment for a newly enrolled
patient, some patients who have not experienced toxicity thus
far may experience toxicity later during the remaining
follow-up. It is worth noting that whether we view toxicity as late-onset
or not is relative to
the patient accrual rate. If patients enter the trial at a fast
rate and toxicity evaluation cannot
keep up with
the speed of enrollment, this situation is considered as late-onset toxicity.
On the other hand, if the patient accrual is very slow, for example, one
patient every three months, and toxicity evaluation also requires a follow-up
of three months, then the trial conduct may not
cause any missing data problem.
For broader applications besides this chemo-radiation trial and to gain
more insight into the missing data issue, we explore several options to
design such late-onset toxicity trials, including the CRM and some
other possibilities discussed below.

Operatively, the CRM does not require that toxicity must be immediately
observable,
and the update of posterior estimates and dose assignment can be
based on the currently observed toxicity data while ignoring the
missing data.
However, such observed data represent a biased sample of the population
because patients who would experience toxicity are more likely
to be included in the sample than
those who do not experience toxicity. In other words,
the observed data contain an excessively higher percentage
of toxicity than the complete data.
Consequently, the estimates based on only the observed data
tend to overestimate the toxicity probabilities
and lead to overly conservative dose escalation.
Alternatively, \citet{CheCha00} proposed
the time-to-event CRM (TITE-CRM),
in which subjects who have not experienced toxicity thus far are
weighted by their follow-up times.
Based on similar weighting methods,
\citet{Bra06} studied both early- and
late-onset toxicities in phase I trials;
Mauguen, Le Deley and Zohar (\citeyear{MauLeDZoh11}) investigated the
EWOC design with incomplete toxicity data; and Wages, Conaway and
O'Quigley (\citeyear{WagConOQu13}) proposed a dose-finding method for
drug-combination trials.
Yuan and Yin (\citeyear{YuYu11}) proposed an expectation--maximization (EM) CRM
approach to handling late-onset toxicity.

In the Bayesian paradigm, we propose a data augmentation
approach to resolving
the late-onset toxicity problem based upon the missing data methodology
[\citet{LitRub02}; and \citet{DanHog08}].
By treating the unobserved toxicity outcomes as missing data,
we naturally integrate the missing data technique and theory into the
CRM framework.
In particular, we establish that the missing data due to late-onset
toxicities are nonignorable.
We propose the Bayesian data augmentation CRM (DA-CRM)
to iteratively impute the missing data and sample from the
posterior distribution of the model parameters based on
the imputed likelihood.

The remainder of the article is organized as follows. In Section~\ref{sec2}
we briefly review the original CRM methodology and
propose the DA-CRM based on Bayesian data augmentation to address
the missing data issue caused by
late-onset toxicity.
In Section~\ref{sec3.1} we present
simulation studies to compare the operating characteristics of the
new design with other available methods, and in Section~\ref{sec3.2}
we conduct a sensitivity analysis
to further investigate the properties of the DA-CRM. We illustrate the
proposed DA-CRM design with two cancer clinical trials in Section~\ref{sec4},
and conclude with a brief discussion in Section~\ref{sec5}.

\section{Dose-finding methods}\label{sec2}
\subsection{Continual reassessment method}\label{sec2.1}
In a phase I dose-finding trial,
patients enter the study sequentially and are
followed for a fixed period of time $(0, T)$ to
assess the toxicity of the drug.
During this evaluation window $(0, T)$, we measure
a binary toxicity outcome for each subject $i$,
\[
Y_i = \cases{1,&\quad $\mbox{if a
drug-related toxicity is observed in } (0, T)$,
\vspace*{2pt}\cr
0,&\quad $\mbox{if no drug-related toxicity is observed in } (0, T).$}
\]
Typically, the length of the assessment period
$T$ is chosen so that if a drug-related toxicity occurs, it would
occur within $(0, T)$.
Depending on the nature of the disease and the treatment
agent, the assessment period $T$ may vary from days to months.

Suppose that a set of $J$
doses of a new drug are under investigation,
the CRM assumes a working dose--toxicity curve, such as
\[
\pi_{d}=\alpha_{d}^{\exp(a)},\qquad d=1, \ldots, J,
\]
where $\pi_{d}$ is the true
toxicity probability at dose level $d$, $\alpha_d$ is the prespecified
probability constant, satisfying a monotonic dose--toxicity order
$\alpha_1<\cdots<\alpha_J$,
and $a$ is an unknown parameter. We
continuously update this dose--toxicity curve by re-estimating $a$
based on the
observed toxicity outcomes in the trial.

Suppose that $n$ patients have entered the trial, and let $y_i$
and $d_i$ denote the binary toxicity outcome and the received dose
level for the $i$th subject, respectively.
The likelihood function based on the toxicity outcomes
$\by=\{y_i, i=1, \ldots, n\}$ is given by
\[
L (\mathbf{y}|a)=\prod_{i=1}^n \bigl
\{\alpha_{d_i}^{\exp(a)} \bigr\} ^{y_i} \bigl\{1-
\alpha_{d_i}^{\exp(a)} \bigr\}^{1-y_i}.
\]
Assuming a prior distribution $f(a)$ for
$a$, for example, $f(a)$ is
a normal distribution with mean 0 and variance $\sigma^2$,
$a\sim N(0, \sigma^2)$, then the posterior distribution of~$a$ is given by
%
%
\begin{equation}
f(a|\by) = \frac{L(\mathbf{y}|a)f(a)} {
\int L(\mathbf{y}|a)f(a) \,\mathrm{d} a}, \label{posterior}
\end{equation}
and the posterior means of the dose--toxicity probabilities are given by
\[
\hat{\pi}_d = \int\alpha_d^{\mathrm{exp}(a)} f(a|\by)\,
\mathrm{d}a, \qquad d=1, \ldots, J.
\]
Based on the updated estimates of the toxicity probabilities, the CRM
assigns a new cohort of patients to dose level $d^*$
which has an estimated toxicity probability closest to the prespecified
target $\phi$,
that is,
\[
d^*=\mathop{\operatorname{argmin}}_{d\in(1,\ldots, J)}| \hat{\pi}_{d}-\phi|.
\]
The trial continues until the exhaustion of the total sample size,
and then the dose with an estimated toxicity probability closest
to $\phi$ is selected as the MTD.

\begin{figure}[b]

\includegraphics{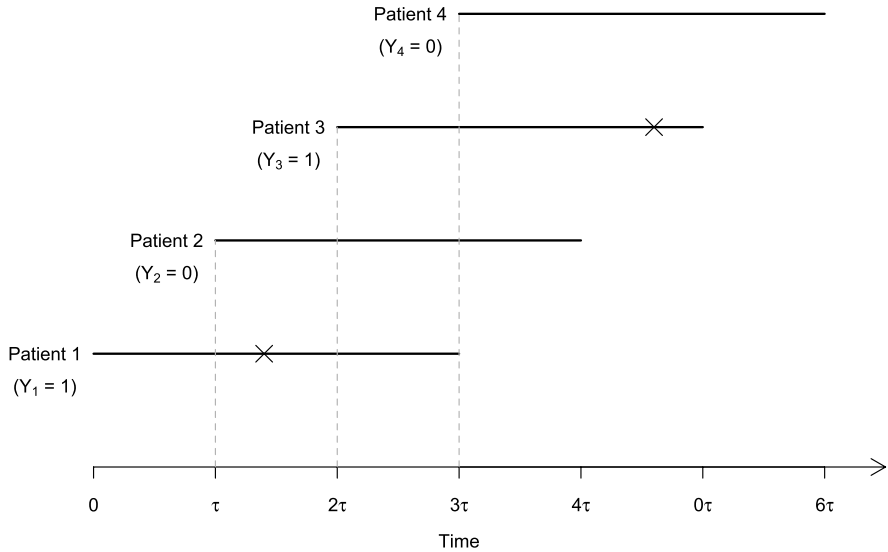}

\caption{Illustration of missing toxicity outcomes under fast accrual.
For each patient,
the horizontal line segment represents the follow-up, on which toxicity is
indicated by a cross. At time $\tau$, the toxicity outcome of patient 1
is missing (i.e., $Y_1$ is missing); at time $2\tau$, the toxicity
outcome of patient 2 is missing (i.e.,
$Y_1=1$, but $Y_2$ is missing); and at time $3\tau$, the toxicity
outcomes of
both patients 2 and~3 are missing (i.e.,
$Y_1=1$, but $Y_2$ and $Y_3$ are missing).}\label{diagram}
\end{figure}

\subsection{Nonignorable missing data}\label{sec2.2}
One of the practical limitations of the CRM is that the DLT
needs to be ascertainable quickly after the initiation of the
treatment. Figure~\ref{diagram} illustrates the situation where
the patient interarrival time $\tau$
is shorter than the assessment period $T$.
By the time a dose is to be assigned to a newly
accrued patient (say, patient 4 at time $3\tau$),
some of the patients who have entered the
trial (i.e., patients 2 and 3) may have been partially
followed and their toxicity outcomes are still not available.
More precisely, for the $i$th subject, let $t_i$ denote the time to
toxicity. For subjects who do not
experience toxicity during the trial, we set $t_i=\infty$. At the
moment of decision making for dose assignment,
let $u_i$ ($0 \le u_i \le T$) denote the actual follow-up time
for subject $i$, and let $M_i(u_i)$ be the missing
data indicator for $Y_i$. Then it follows that
%
%
\begin{equation} \label{missmech}
M_i(u_i) = \cases{ %
\mathrm{1},
&\quad $\mbox{if } t_i>u_i \mbox{ and }
u_i<T$,
\vspace*{2pt}\cr
0,&\quad $\mbox{if } t_i \le u_i \mbox{ or } u_i=T.$}
\end{equation}
That is, the toxicity outcome is
missing with $M_i(u_i)=1$ for patients who
have not yet experienced toxicity ($t_i>u_i$) and have not
been fully followed up to $T$ ($u_i<T$); and the toxicity outcome is
observed with $M_i(u_i)=0$ when patients either have experienced toxicity
($t_i \le u_i$) or have completed the entire
follow-up ($u_i=T$) without experiencing toxicity.
For notational simplicity, we suppress $u_i$ and take
$M_i\equiv M_i(u_i)$.
Due to patients' staggered entry, it is reasonable
to assume that $u_i$ is independent
of $t_i$, that is, the time of dose assignment (or the arrival of a
new patient) is independent of the time to toxicity.

Under the missing data mechanism (\ref{missmech}),
the induced missing data are nonignorable or informative
because the probability of missingness of
$Y_i$ depends on the underlying time to toxicity, and thus
implicitly depends on the value of $Y_i$ itself. More
specifically, the data from patients who would not experience
toxicity ($Y_i=0$) in the assessment period are more likely to be missing
than data from patients who would experience toxicity ($Y_i=1$).
The next theorem provides a new insight to the issue
of late-onset toxicity.

\begin{theorem}\label{th1} Under the missing data mechanism (\ref{missmech}),
the missing data induced by late-onset toxicity
are nonignorable with $\Pr(M_i=1|Y_i=0) > \Pr(M_i=1|Y_i=1)$.
\end{theorem}

The proof of the theorem is briefly sketched in the \hyperref[app]{Appendix}.
In general, the missing data are more likely to occur for
those patients who would not experience toxicity in $(0, T)$.
This phenomenon is also illustrated in Figure~\ref{diagram}. Patient~2
who will not experience toxicity during the assessment period
is more likely to have a missing toxicity outcome at the
decision-making times
$2\tau$ and $3\tau$ than patient~1 who has experienced toxicity
between times $\tau$ and $2\tau$.
Compared with other missing data mechanisms, such as
missing completely at random or
missing at random, nonignorable missing data are the most
difficult to deal with [\citet{LitRub02}],
which brings a new challenge to clinical trial designs.
Because the missing data are nonignorable, the naive approach by simply
discarding the missing data and making dose-escalation decisions solely
based on
the observed toxicity data is problematic.
The observed data represent a biased sample of the
complete data and contain more toxicity observations than they should be
because the responses for patients who would experience toxicity are more
likely to be observed. As a result, approaches based only on the observed
toxicity data typically overestimate the toxicity probabilities and
thus lead
to overly conservative dose escalation.

During the trial conduct, the amount of missing data depends on the
ratio of the assessment period $T$ and the interarrival time $\tau$,
denoted as
the A/I ratio $=T/\tau$. The larger the value of the
A/I ratio, the greater the amount of missing data that would be
produced, because there would be
more patients who may not have completed the toxicity assessment
when a new cohort arrives.

\subsection{DA-CRM using Bayesian data augmentation}\label{sec2.3}
An intuitive approach to dealing with the unobserved toxicity
outcomes is to impute the missing data so that the standard
complete-data method can be applied.
One way to achieve this goal is to use data augmentation (DA) proposed
by \citet{TanWon87}.
The DA iterates between two steps: the imputation (I)
step, in which the missing data are imputed, and the posterior (P) step,
in which the posterior samples of unknown parameters are simulated
based on
the imputed data.
As the CRM is originally formulated in
the Bayesian framework [O'Quigley, Pepe and
Fisher (\citeyear{OQuPepFis90})], the
DA provides a natural
and coherent way to address the missing data
issue due to late-onset toxicity. Note that the missing data we
consider here is a special
case of nonignorable missing data
with a known missing data mechanism as defined by~(\ref{missmech}).
Therefore, the nonidentification problem that often plagues
the nonignorable missing data
can be circumvented as follows.

In order to obtain consistent estimates, we need to model the
nonignorable missing
data mechanism in (\ref{missmech}). Toward this goal, we specify a
flexible piecewise
exponential model for the time to toxicity for patients who would
experience DLTs, which
concerns the conditional distribution of $t_i$ given $Y_i=1$. Specifically,
we consider a partition of the follow-up period $[0, T]$ into
$K$ disjoint intervals $[0, h_1), [h_1, h_2), \ldots, [h_{K-1},
h_K\equiv T]$,
and assume a constant hazard $\lambda_k$
in the $k$th interval. Define the observed time $x_i=\mathrm{min}(u_i, t_i)$
and $\delta_{ik}=1$ if the $i$th subject experiences toxicity in the
$k$th interval;
and $\delta_{ik}=0$ otherwise. Let
$\blambda= \{\lambda_1, \ldots, \lambda_K\}$; when
the toxicity data $\by=\{y_1,\ldots,y_n\}$ are completely observed,
the likelihood function of $\blambda$ based on
$n$ enrolled subjects is given by
\[
L(\by|\blambda)=\prod_{i=1}^n\prod
_{k=1}^K \lambda_k^{\delta_{ik}}
\operatorname{exp}\{-y_i\lambda_k s_{ik}\},
\]
where $s_{ik}=h_k-h_{k-1}$ if $x_i>h_k$; $s_{ik}=x_i-h_{k-1}$ if
$x_i\in[h_{k-1}, h_k)$;
and otherwise $s_{ik}=0$. Similar to the TITE-CRM, we
assume that the time-to-DLT distribution is invariant to
the dose level, conditioning on that the patient will experience toxicity
($Y_i=1$). This assumption is helpful to pool information across
different doses and obtain more reliable estimates. The sensitivity
analysis in
Section~\ref{sec3.2} shows that our method is not sensitive to
the violation of this assumption.

In the Bayesian paradigm, we assign each component of $\blambda$ an
independent gamma prior distribution with the shape parameter $\zeta_k$
and the rate parameter $\xi_k$,
denoted as $\operatorname{Ga}(\zeta_k, \xi_k)$. When there is some prior knowledge
available regarding the shape of the hazard for
the time to toxicity, the hyperparameters $\zeta_k$ and $\xi_k$ can
be calibrated to match the prior information.
Here we focus on the common case in which the prior information is
vague and
we aim to develop a default and automatic prior
distribution for general use. Specifically,
we assume that {a priori} toxicity occurs uniformly throughout
the assessment period $(0, T)$,
which represents a neutral prior opinion between early-onset and
late-onset toxicity. Under this assumption,
the hazard at the middle of the $k$th partition is $\tilde{\lambda
}_k=K/\{T(K-k+0.5)\}$.
Thus, we assign $\lambda_k$ a gamma prior distribution,
\[
\lambda_k = \operatorname{Ga}(\tilde{\lambda}_k/C, 1/C),
\]
where $C$ is a constant determining the size of the variance with
respect to the mean, as the mean for
this prior distribution is $\tilde{\lambda}_k$ and the variance is
$C\tilde{\lambda}_k$.
Based on our simulations, we
found that $C = 2$ yields a reasonably vague prior and
equips the design with good operating characteristics.

Based on the time-to-toxicity model as above, the DA algorithm can be
implemented as follows.
At the I step of the DA, we ``impute'' the missing data by drawing
posterior samples from their
full conditional distributions. Let $\by=(\by_{\mathrm{obs}}, \by_{\mathrm{mis}})$,
where $\by_{\mathrm{obs}}$ and $\by_{\mathrm{mis}}$
denote the observed and missing toxicity data, respectively; and
let $\cal{D}_\mathrm{obs} = (\by_\mathrm{obs}, \bM)$ denote the
observed data
with missing indicators $\bM=\{M_i, i=1, \ldots, n\}$.
As the missing data are informative, the
observed data used for inference include
not only the observed toxicity outcomes $\by_\mathrm{obs}$,
but also the missing data indicators $\bM$. Inference that ignores $\bM
$ (such as the CRM) would
lead to biased estimates. It can be shown that, conditional on the
observed data
$\cal{D}_\mathrm{obs}$ and model parameters $(a, \blambda)$,
the full conditional distribution of $y_i \in\by_\mathrm{mis}$ is
given by
\[
y_{i} | ({\cal D}_\mathrm{obs},a, \blambda) \sim \operatorname{Bernoulli} \biggl(\frac{ \alpha_{d_i}^{\mathrm{exp}{({a})}} \operatorname{exp}(-\sum_{k=1}^K \lambda_ks_{ik}) } {
1-\alpha_{d_i}^{\mathrm{exp}{({a})}} + \alpha_{d_i}^{\mathrm{exp}{({a})}}
\operatorname{exp}(-\sum_{k=1}^K \lambda_ks_{ik})} \biggr).
\]

At the P step of the DA, given the imputed data $\by$,
we sequentially sample the unknown model parameters from their
full conditional distributions as follows:
\begin{longlist}[(ii)]
\item[(i)] Sample $a$ from $f(a| \by)$
given by (\ref{posterior}), where $\by$ is the ``complete''
data after filling in the missing outcomes.
\item[(ii)] Sample $\lambda_k, k=1, \ldots, K$, from
\[
\lambda_k|\by\sim \operatorname{Ga} \Biggl(\tilde{\lambda}_k/C+\sum
_{i=1}^n{\delta_{ik}}, 1/C+\sum
_{i=1}^n y_is_{ik}
\Biggr).
\]
\end{longlist}
The DA procedure iteratively draws a sequence of samples of the missing
data and
model parameters through the imputation (I) step and posterior (P) step
until the Markov chain converges.
The posterior samples of $a$
can then be used to make inference on $\pi_d$ to direct dose finding.

\subsection{Dose-finding algorithm}\label{sec2.4}
Let ${\phi}$ denote the physician-specified toxicity target, and assume that
patients are treated in cohorts, for example, with a cohort size of
three. For safety, we restrict dose escalation or
de-escalation by one dose level of change at a time. The dose-finding
algorithm for the DA-CRM is described as follows:
\begin{longlist}[(1)]
\item[(1)] Patients in the first cohort are treated at
the lowest dose level.
\item[(2)]
At the current dose level $d^\mathrm{curr}$, based on the cumulated data,
we obtain the posterior means
for the toxicity probabilities, $\hat{\pi}_d\ (d=1, \ldots, J)$.
We then find dose level $d^*$ that has a toxicity probability closest to
$\phi$, that is,
\[
d^*=\mathop{\operatorname{argmin}}_{d\in(1,\ldots, J)}| \hat{\pi}_{d}-\phi|.
\]
\begin{itemize}
\item
If $d^\mathrm{curr} > d^*$, we de-escalate the dose level to $d^\mathrm
{curr}-1$;
\item if $d^\mathrm{curr} < d^*$, we escalate the dose level to
$d^\mathrm{curr}+1$;
\item otherwise, the dose stays at the same level as
$d^\mathrm{curr}$ for the next cohort of patients.
\end{itemize}

\item[(3)] Once the maximum sample size is reached,
the dose that has the toxicity probability closest to $\phi$ is
selected as the MTD.
\end{longlist}
In addition, we also impose an early stopping rule for safety:
if $\Pr(\pi_1> \phi|\cD)>0.96$, the trial will be terminated.
That is, if the lowest dose is still overly toxic, the trial
should be stopped early.

\section{Numerical studies}\label{sec3}
\subsection{Simulations}\label{sec3.1}
To examine the operating characteristics of the DA-CRM design, we
conducted extensive simulation studies. We considered six
dose levels and assumed that toxicity monotonically
increased with respect to the dose. The target toxicity
probability was 30\% and a maximum number of 12 cohorts were
treated sequentially in a cohort size of three. The sample size was
chosen to match the maximum sample size required by the conventional
``3${}+{}$3'' design. The toxicity assessment period was $T=3$ months and the
accrual rate was 6 patients per month. That is, the interarrival time
between every two consecutive cohorts was $\tau=0.5$
month with the A/I ratio${} = {}$6.

We considered four toxicity scenarios
in which the MTD was located at different dose levels. Due to the
limitation of space, we show only scenarios 1 and 2 in Table~\ref{tab1}, and the
other scenarios are provided in Table S1 of the supplementary materials [\citet{supp}].
Under each scenario, we simulated times to toxicity based on
Weibull, log-logistic and uniform distributions,
respectively. For Weibull and log-logistic distributions,
we controlled that 70\% toxicity events would occur
in the latter half of the assessment period $(T/2, T)$.
Specifically, at each dose level, the scale and shape
parameters of the Weibull distribution were chosen such that
\begin{longlist}[(1)]
\item[(1)] the cumulative distribution function at the end of the
follow-up time $T$
would be the toxicity probability of that dose; and
\item[(2)]
among all the toxicities that occurred in $(0, T)$, 70\% of them would
occur in $(T/2, T)$,
the latter half of the assessment period.
\end{longlist}
Because the toxicity probability varies across
different dose levels, the scale and shape
parameters of the Weibull distribution need to be carefully chosen for
different dose levels, and similarly for the scale and location
parameters of the
log-logistic distribution. For the uniform distribution, we simulated
the time to toxicity independently for each dose level and controlled
the cumulative \mbox{distribution} function at the end of the follow-up time
$T$ matching the toxicity probability of each dose.
In the proposed DA-CRM, we used $K=9$ partitions to construct the piecewise
exponential time-to-toxicity model.
We compared the DA-CRM with the CRM$_\mathrm{obs}$, which determined
the dose assignment
based on only the observed toxicity data as suggested by O'Quigley, Pepe and
Fisher (\citeyear{OQuPepFis90}),
and the TITE-CRM with the adaptive weighting scheme proposed by
\citet{CheCha00}.
As a benchmark for comparison, we also implemented the complete-data version
of the CRM (denoted by CRM$_\mathrm{comp}$), assuming that
all of the toxicity outcomes in the trial were completely observed
prior to
each dose assignment. The CRM$_\mathrm{comp}$ required repeatedly
suspending the accrual prior to each dose assignment to wait that all
of the toxicity
outcomes in the trial were completely observed.
Although the CRM$_\mathrm{comp}$ is not feasible in practice when toxicities
are late-onset, it provides an optimal upper bound to evaluate the
performances of other designs. Actually, when all toxicity outcomes are
observable (i.e., no missing data),
the DA-CRM and TITE-CRM are equivalent to the complete-data CRM$_\mathrm{comp}$.
For all methods, we set the probability constants
in the CRM $(\alpha_1, \ldots, \alpha_6)=(0.08, 0.12, 0.20, 0.30, 0.40,
0.50)$ and used
a normal prior distribution $N(0,2)$ for parameter $a$.
Under each scenario, we simulated 5000 trials.

\begin{table}
\tabcolsep=0pt
\caption{Simulation study comparing the complete-data CRM (CRM$_\mathrm{comp}$),
the CRM based on the observed toxicity data only (CRM$_\mathrm{obs}$),
time-to-event CRM
(TITE-CRM) and the proposed data augmentation CRM (DA-CRM) with the
sample size
36, the cohort size 3 and the A/I ratio $=6$}\label{tab1}
\fontsize{8.8}{10.4}{\selectfont{
\begin{tabular*}{\textwidth}{@{\extracolsep{4in minus 4
in}}lcd{2.2}d{2.2}d{2.1}d{2.2}d{2.2}d{1.1}d{2.2}cd{2.1}@{}}
\hline
&&\multicolumn{7}{c}{\textbf{Recommendation percentage at dose level}}
& &
\\[-6pt]
&&\multicolumn{7}{c}{\hrulefill} & & \\
\multicolumn{1}{@{}l}{\multirow{2}{35pt}[10pt]{\textbf{Time to
toxicity}}} &
\multicolumn{1}{c}{\textbf{Design}} & \multicolumn{1}{c}{\textbf{1}} &
\multicolumn{1}{c}{\textbf{2}} &
\multicolumn{1}{c}{\textbf{3}} & \multicolumn{1}{c}{\textbf{4}} &
\multicolumn{1}{c}{\textbf{5}} & \multicolumn{1}{c}{\textbf{6}} &
\multicolumn{1}{c}{\textbf{None}} &
\multicolumn{1}{c}{$\bolds{N_\mathrm{MTD+}}$} &\multicolumn
{1}{c@{}}{\multirow{2}{35pt}[10pt]{\textbf{Duration (months)}}} \\
\hline
Scenario 1 & Pr(toxicity) & 0.1 & 0.15 & 0.3 & 0.45 & 0.6 & 0.7 & & \\
& CRM$_\mathrm{comp}$ & 0.6 & 13.8 & \multicolumn{1}{c}{\textbf{61.9}}
& 22.9 & 0.6 & 0.0 & 0.2 & \phantom{0}\textbf{9.0} & 36.4 \\
& \# patients & 4.8 & 7.2 & 14.9 & 7.6 & 1.3 & 0.1 & & \\[3pt]
Weibull & CRM$_\mathrm{obs}$& 0.4 & 7.5 & \multicolumn{1}{c}{\textbf
{48.4}} & 27.4 & 2.4 & 0.1 & 13.7 & \phantom{0}\textbf{4.0} & 8.2 \\
& \# patients & 7.5 & 8.3 & 13.2 & 3.0 & 0.9 & 0.1 & & \\
& TITE-CRM & 3.4 & 23.1 & \multicolumn{1}{c}{\textbf{55.9}} & 16.5 &
0.6 & 0.0 & 0.5 & \textbf{15.5} & 9.0 \\
& \# patients & 5.1 & 6.3 & 9.0 & 8.1 & 4.9 & 2.5 & & \\
& DA-CRM & 0.9 & 14.7 & \multicolumn{1}{c}{\textbf{56.4}} & 25.1 & 1.5
& 0.0 & 1.2 & \textbf{10.4} & 8.9 \\
& \# patients & 9.4 & 7.6 & 8.3 & 6.0 & 3.0 & 1.3 & & \\[3pt]
Log-logistic &CRM$_\mathrm{obs}$& 0.3 & 7.8 &\multicolumn{1}{c}{\textbf
{48.2}} & 27.4 & 2.7 & 0.1 & 13.6 & \phantom{0}\textbf{4.0} & 8.2 \\
& \# patients & 7.4 & 8.4 & 13.2 & 3.0 & 0.9 & 0.1 & & \\
&TITE-CRM & 3.6 & 22.6 & \multicolumn{1}{c}{\textbf{56.3}} & 16.5 & 0.6
& 0.0 & 0.4 & \textbf{15.4} & 9.0 \\
& \# patients & 5.1 & 6.4 & 9.1 & 8.2 & 4.8 & 2.4 & & \\
& DA-CRM & 0.9 & 13.9 & \multicolumn{1}{c}{\textbf{58.1}} & 23.8 & 1.9
& 0.0 & 1.3 & \textbf{10.3} & 8.9 \\
& \# patients & 9.5 & 7.6 & 8.2 & 6.0 & 3.0 & 1.3 & &\\[3pt]
Uniform &CRM$_\mathrm{obs}$& 0.1 & 4.7 & \multicolumn{1}{c}{\textbf
{38.0}} & 30.5 & 2.9 & 0.1 & 23.6 & \phantom{0}\textbf{2.8} & 7.5 \\
& \# patients & 8.6 & 8.2 & 10.9 & 2.3 & 0.5 & 0.0 & & \\
&TITE-CRM & 2.1 & 20.4 & \multicolumn{1}{c}{\textbf{56.6}} & 19.1 & 1.4
& 0.0 & 0.4 & \textbf{13.0} & 9.0 \\
& \# patients & 6.3 & 7.0 & 9.7 & 8.0 & 3.7 & 1.3 & & \\
& DA-CRM & 0.6 & 11.2 & \multicolumn{1}{c}{\textbf{56.9}} & 27.6 & 1.7
& 0.0 & 1.9 & \phantom{0}\textbf{8.7} & 8.9 \\
& \# patients & 10.8 & 7.6 & 8.5 & 5.7 & 2.2 & 0.7 & & \\[6pt]
Scenario 2 & Pr(toxicity) & 0.08 & 0.1 & 0.2 & 0.3 & 0.45 & 0.6 & & \\
& CRM$_\mathrm{comp}$ & 0.0 & 1.4 & 23 & \multicolumn{1}{c}{\textbf
{55.9}\phantom{0}} & 18.8 & 0.8 & 0.1 & \phantom{0}\textbf{6.6} & 36.4
\\
& \# patients & 4.1 & 4.1 & 9.0 & 12.2 & 5.5 & 1.0 & &\\[3pt]
Weibull &CRM$_\mathrm{obs}$& 0.0 & 1.0 & 18.9 & \multicolumn
{1}{c}{\textbf{48.5}\phantom{0}} & 22.2 & 1.8 & 7.6 & \phantom{0}\textbf
{2.9} & 8.5 \\
& \# patients & 5.7 & 6.7 & 13.8 & 5.2 & 2.3 & 0.7 & & \\
&TITE-CRM & 0.1 & 2.6 & 29.2 & \multicolumn{1}{c}{\textbf{52.4}\phantom
{0}} & 14.9 & 0.8 & 0.1 & \textbf{11.2} & 9.0 \\
& \# patients & 4.2 & 4.7 & 7.1 & 8.8 & 6.8 & 4.4 & & & \\
& DA-CRM & 0.0 & 1.5 & 24.4 & \multicolumn{1}{c}{\textbf{54.0}\phantom
{0}} & 17.7 & 1.4 & 1.1 & \phantom{0}\textbf{7.3} & 8.9 \\
& \# patients & 8.4 & 6.3 & 7.0 & 6.7 & 4.5 & 2.8 & & & \\[3pt]
Log-logistic &CRM$_\mathrm{obs}$& 0.0 & 0.9 & 19.0 & \multicolumn
{1}{c}{\textbf{48.5}\phantom{0}} & 22.0 & 1.9 & 7.7 & \phantom{0}\textbf
{2.9} & 8.5 \\
& \# patients & 5.7 & 6.7 & 13.8 & 5.1 & 2.3 & 0.7 & & \\
& TITE-CRM & 0.1 & 2.6 & 29.0 & \multicolumn{1}{c}{\textbf{52.3}\phantom
{0}} & 15.1 & 0.8 & 0.1 & \textbf{11.1} & 9.0 \\
& \# patients & 4.1 & 4.7 & 7.2 & 8.8 & 6.8 & 4.3 & & & \\
& DA-CRM & 0.0 & 1.4 & 23.3 & \multicolumn{1}{c}{\textbf{54.0}\phantom
{0}} & 18.7 & 1.6 & 1.0 & \phantom{0}\textbf{7.5} & 8.9 \\
& \# patients & 8.3 & 6.2 & 6.9 & 6.8 & 4.6 & 2.9 & & & \\[3pt]
Uniform &CRM$_\mathrm{obs}$& 0.0 & 0.7 & 14.9 & \multicolumn
{1}{c}{\textbf{45.3}\phantom{0}} & 23.3 & 2.3 & 13.5 & \phantom
{0}\textbf{2.0} & 8.1 \\
& \# patients & 6.9 & 7.1 & 12.3 & 4.5 & 1.7 & 0.4 & & \\
& TITE-CRM & 0.1 & 2.2 & 26.9 & \multicolumn{1}{c}{\textbf{54.4}\phantom
{0}} & 15.4 & 0.9 & 0.1 & \phantom{0}\textbf{9.6} & 9.0 \\
& \# patients & 4.9 & 5.0 & 7.6 & 8.8 & 6.2 & 3.3 & & & \\
& DA-CRM & 0.0 & 1.3 & 23.7 & \multicolumn{1}{c}{\textbf{54.0}\phantom
{0}} & 18.8 & 1.1 & 1.2 & \phantom{0}\textbf{6.2} & 8.9 \\
& \# patients & 9.3 & 6.3 & 7.1 & 6.8 & 4.2 & 2.0 & & & \\
\hline
\end{tabular*}
}}
\end{table}

Following each scenario in Tables~\ref{tab1} and S1,
the first row is the true toxicity probabilities; rows
2 and 3 show the dose selection probability (with the percentage
of inconclusive trials denoted by ``None'') and the average number
of patients treated at each dose based on the complete-data
design CRM$_\mathrm{comp}$, respectively; the remaining rows provide the
corresponding summary statistics
for the CRM$_\mathrm{obs}$, TITE-CRM and DA-CRM under various settings of
late-onset toxicity and time-to-toxicity distributions.
The CRM$_\mathrm{comp}$ does not depend on the distributions of the
times to toxicity because the design assumes that all toxicity outcomes
are completely observed before each dose assignment.

When evaluating the trial designs with late-onset toxicities, one of
the most important measures of the design performance is patient
safety because the main issue of the late-onset toxicities is that
ignoring them will lead to overly aggressive dose escalation and
thus treating too many patients at excessively toxic doses,
that is, the doses higher than the MTD.
As a measure of safety, in Tables~\ref{tab1} and S1,
we also report the number of patients treated at doses above the MTD
(denoted as
$N_\mathrm{MTD+}$) averaged across 5000
simulated trials.

In scenario 1, the MTD (shown in boldface) is at dose level 3,
and the complete-data design CRM$_\mathrm{comp}$ yielded an optimal
selection probability of 61.9\%. The selection probability of the MTD using
the DA-CRM was slightly lower than this optimal value, but higher than that
of using the CRM$_\mathrm{obs}$.
For instance, when the time to toxicity followed the log-logistic
distribution, the selection probability
using the DA-CRM was 58.1\%, whereas that of the
CRM$_\mathrm{obs}$ was 48.2\%. The CRM$_\mathrm{obs}$ appeared to be overly
conservative and led to a high percentage (about 13.6\%) of
inconclusive trials.
This was because the CRM$_\mathrm{obs}$ estimated the toxicity probabilities
based solely on the observed toxicity data, which is
a biased sample of the complete data with an excessive number of toxicities.
Therefore, the CRM$_\mathrm{obs}$ tended to overestimate the toxicity
probabilities, resulting in conservative dose escalations and high
percentages of early termination of the trial.
The TITE-CRM yielded similar selection percentages as the DA-CRM, but
the DA-CRM was much safer: the number of patients treated above the MTD
(i.e., $N_\mathrm{MTD+}$) using the DA-CRM was notably smaller than that
of the TITE-CRM and close to that of the complete-data design. For
example, when the time to toxicity followed the Weibull distribution,
$N_\mathrm{MTD+}$ was 9.0 and 10.4
using the complete-data design and the DA-CRM, respectively, while that
based on the TITE-CRM was 15.5. As the CRM$_\mathrm{obs}$ is overly
conservative, $N_\mathrm{MTD+}=4.0$ is the smallest under the
CRM$_\mathrm{obs}$.

In scenario 2, the MTD is the fourth dose, and in scenario 3 (see Table S1),
the MTD is the second.
Compared to the TITE-CRM, the DA-CRM yielded comparable
MTD-selection probabilities but appeared
to be safer, which reduced $N_\mathrm{MTD+}$
by more than 30\% in both scenarios. For example, in scenario 2, when
the time to toxicity followed the Weibull distribution,
$N_\mathrm{MTD+}$ using the DA-CRM was 7.3, approximately 35\% less
than that
using the TITE-CRM ($N_\mathrm{MTD+}=11.2$). A similar
extent of decreasing in $N_\mathrm{MTD+}$ was
observed in scenario 3 when using the DA-CRM. The
CRM$_\mathrm{obs}$ again led to a high percentage of inconclusive trials
(particularly under the uniform distribution)
and a relatively low selection percentage
of the MTD due to the overestimation of the toxicity probabilities.
For scenario~4 in Table S1, in which the fifth dose is the MTD,
the CRM$_\mathrm{obs}$ yielded a similar selection percentage as the
TITE-CRM and DA-CRM.

We further investigated the performance of the designs under a
smaller sample size of 27 patients treated in a cohort size of 3,
and 21 patients treated in a cohort size of 1. The pattern of
the results is generally similar to those described above
(see Tables S2 and S3 in the supplementary materials [\citet{supp}]). We also examined
the operating characteristics of the DA-CRM under a lower A/I
ratio of 3 with the cohort interarrival time $\tau=1$ month
(see Table S4 in the supplementary materials [\citet{supp}]).
In this case, the accrual rate was relatively slower and,
thus, late-onset toxicities became
of less concern since the majority of toxicity outcomes would be
observed at the moment of dose assignment. As expected,
the performances of the CRM$_\mathrm{obs}$, TITE-CRM and DA-CRM were rather
comparable across different scenarios and time-toxicity distributions.
Actually, when the A/I ratio is less than or equal to 1 (i.e., no
late-onset toxicities
and no missing data), the CRM$_\mathrm{obs}$, TITE-CRM and DA-CRM are exactly
the same.

These results suggest that, when the A/I ratio is low (e.g., when the
disease under study is rare and thus the accrual rate is slow), the
CRM$_\mathrm{obs}$ has little bias and is still a good design option
for phase I clinical trials. However, when the accrual is fast, for
example, in multi-center clinical trials for
some common type of cancer (e.g., breast or lung cancer), the A/I ratio
can be high (particularly when radiotherapies or some targeted agents
are used), and using the proposed DA-CRM can lead to better operating
characteristics.

\subsection{Sensitivity analysis}\label{sec3.2}

We investigated the robustness of the proposed
DA-CRM design when (1) the underlying
times to toxicity were heterogeneous across the doses,
by simulating the times to toxicity from a Weibull distribution at
dose levels of 1, 3 and 5, and from a log-logistic distribution
at dose levels of 2, 4 and 6; (2) the number of partitions
used in the piecewise exponential model for the times
to toxicity was $K=5$ and 12; and (3) the prior distribution for $a$ was
$N(0, 0.57)$, the ``least-informative'' prior proposed by \citet
{LeeChe11}. The results show that the performance of the DA-CRM (e.g.,
the selection percentages and $N_\mathrm{MTD+}$) was very similar across
different conditions (see Table S5 in the supplementary materials [\citet{supp}]),
which suggests the robustness of the proposed design.

\section{Applications}\label{sec4}

\subsection{Pancreatic cancer trial}\label{sec4.1}
Muler et~al. (\citeyear{MulMcGNor}) described a phase I trial to
determine the MTD of cisplatin that could be added to
the full-dose gemcitabine and radiation therapy in patients with
pancreatic cancer.
The protocol treatment consisted of two 28-day cycles of chemotherapy,
with radiation given during the first cycle of chemotherapy.
Radiation and gemcitabine doses were held constant, while four dose levels
of cisplatin (20, 30, 40 and 50 mg/m$^2$) were investigated in the trial.
The DLTs were defined as CTCAE 2.0 grade 4 thrombocytopenia,
grade 4 neutropenia lasting more than 7 days or grade 3 toxicity in
other organ systems. Patients were required to be followed for
nine weeks in order to fully assess their toxicity outcomes.
The goal of the trial was to determine the dose of cisplatin associated
with a target
DLT rate of 20\%.

As shown in Table~\ref{app2data}, one challenge of designing this trial
is that the accrual was fast, compared to the 9-week assessment period
for DLTs (i.e., the toxicity was late-onset). In the DA-CRM design, we
took $\alpha=(0.1, 0.15,\break  0.2, 0.25)$ as the prior estimates of the
toxicity probabilities for the four dose levels of cisplatin, and used
30 mg/m$^2$ as the starting dose
of the trial. For patient safety, we required that at least two
patients must have fully completed their toxicity assessment at the
lower dose before the dose can be escalated to the next higher level.

\begin{table}
\tabcolsep=0pt
\caption{Days, doses and DLTs for eighteen evaluable patients enrolled
in the pancreatic cancer trial}\label{app2data}
\begin{tabular*}{\textwidth}{@{\extracolsep{\fill}}lcccc@{\qquad\qquad}lcccc@{}}
\hline
\textbf{Patient}& \textbf{Day on} & \textbf{Day off} & \textbf{Dose} &
& \textbf{Patient}& \textbf{Day on} & \textbf{Day off} & \textbf{Dose}
\\
\textbf{No.} & \multicolumn{1}{c}{\textbf{study}$^{\bolds{*}}$} &
\multicolumn{1}{c}{\textbf{study}$^{\bolds{*}}$} &
\multicolumn{1}{c}{\textbf{(mg/m}$^{\bolds{2}}$\textbf{)}} & \textbf
{DLT} &\textbf{No.} &
\multicolumn{1}{c}{\textbf{study}$^{\bolds{*}}$} &
\multicolumn{1}{c}{\textbf{study}$^{\bolds{*}}$} & \multicolumn
{1}{c}{\textbf{(mg/m}$^{\bolds{2}}$\textbf{)}} & \textbf{DLT} \\
\hline
1 & \phantom{00}0 & \phantom{0}67 & 30 & No & 10 & 224 & 291 & 50 & No\\
2 & \phantom{0}43 & \phantom{0}98 & 30 & No & 11 & 280 & 303 & 50 &
Yes\\
3 & \phantom{0}50 & 116 & 30 & No & 12 & 301 & 347 & 50 & Yes\\
4 & \phantom{0}56 & 108 & 30 & No & 13 & 322 & 382 & 50 & No\\
5 & \phantom{0}70 & 133 & 40 & No & 14 & 329 & 389 & 50 & No\\
6 & 147 & 217 & 40 & No & 15 & 343 & 372 & 50 & Yes\\
7 & 161 & 224 & 40 & No & 16 & 364 & 423 & 40 & No\\
8 & 182 & 238 & 40 & No & 17 & 371 & 408 & 50 & Yes\\
9 & 224 & 284 & 50 & No & 18 & 455 & 528 & 30 & No\\
\hline
\end{tabular*}
\tabnotetext[]{}{* days since the initiation of the study.}
\end{table}

\begin{figure}

\includegraphics{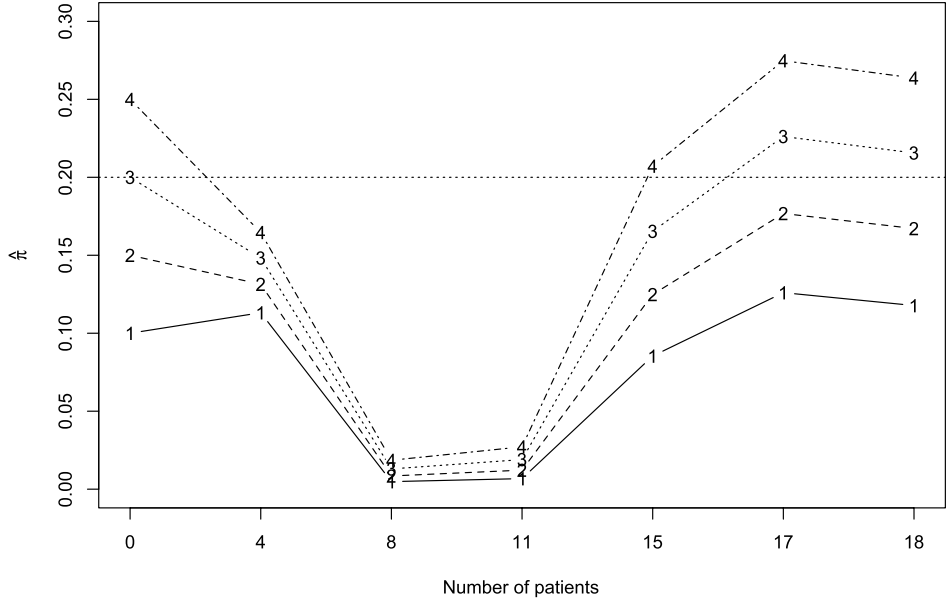}

\caption{Estimates of the toxicity probabilities of four doses
with the cumulative number of patients in the
pancreatic cancer trial. Numbers 1--4 in the figure indicate the
four dose levels.}\label{app2fig}
\end{figure}

Figure~\ref{app2fig} summarizes how the posterior estimates
of the toxicity probabilities of four doses were updated as more patients
were enrolled during the trial conduct.
The first four patients were assigned to the dose of 30 mg/m$^2$.
Based on the days from the initiation
of the trial, when patient 5 arrived on day 70, patient 1 had
completed the follow-up, while patients 2, 3 and 4 had finished
only 43\%, 32\% and 22\% of their follow-ups,
without experiencing any DLTs. The estimates of toxicity probabilities
of four doses were $\hat{\pi}= (0.113, 0.131, 0.148, 0.165)$. We
escalated the dose and subsequently treated patients 5 to 8 at the dose
of 40 mg/m$^2$.
Patient 2 died after 63 days on therapy, but was judged to be secondary
to the
hypercoaguable state associated with pancreatic cancer. Therefore, that
death was classified as unrelated to therapy (i.e., not a DLT).

Upon the arrival of patient 9 on day 224, patients 1 to 7 had completed
their toxicity assessment and none of them had experienced DLT. These
data yielded the updated toxicity estimates $\hat{\pi}= (0.005, 0.008,
0.013, 0.019)$, suggesting that the doses of 30 mg/m$^2$ and 40 mg/m$^2$
were safe. As a result, we further escalated the dose and assigned
patients 9 to 11 to 50 mg/m$^2$. On day 301 when patient 12 was
accrued, patients 1 to 10 had completed their toxicity assessment
without experiencing DLT, yielding the updated estimates of the
toxicity probabilities $\hat{\pi}= (0.007, 0.012, 0.019, 0.027)$.
Consequently, patients 12 to 15 were also treated at 50 mg/m$^2$.

After patient 12 experienced a DLT (i.e., duodenal ulcer) on day 347,
the estimates of
the toxicity probabilities began to increase, that is,
$\hat{\pi}= (0.085, 0.125,  0.165, 0.207)$, but not sufficiently to trigger
dose de-escalation. According to the dose-finding algorithm, the
incoming patients 16 and 17 should be treated at 50~mg/m$^2$. However,
because the
investigators were concerned about a potential DLT in patient 15,
only patient 17 was treated at 50 mg/m$^2$, while patient 16 was treated
at a lower dose 40 mg/m$^2$.

By the time that patient 18, the last enrolled patient, arrived on day 455,
4 out of 8 patients previously treated at 50 mg/m$^2$ had experienced
DLTs (i.e., one duodenal ulcer, one diarrhea resulting in dehydration,
and two grade 3 anorexia and nausea leading to a two-level
decline in performance status). This significantly increased $\hat{\pi
}$ to (0.126, 0.177, 0.228, 0.275). Therefore, patient 18 was assigned
to a lower dose 30 mg/m$^2$. At the end of the trial, the estimates of
the toxicity probabilities were $\hat{\pi}= (0.118, 0.167, 0.215,
0.264)$ and,
thus, the dose 40 mg/m$^2$ was selected as the MTD because
its estimated toxicity probability was closest to the target of~0.2.

\subsection{Esophageal cancer trial}\label{sec4.2}
In the esophageal cancer clinical trial described in \hyperref[secintro]{Introduction},
the target toxicity probability was 30\%
and a total of 30 patients were treated sequentially in cohorts
of size 3. Six doses were investigated and
the trial started by treating the first cohort at dose level 1.
Under the DA-CRM design, the posterior estimates of the dose--toxicity
probabilities
were updated only when the first patient of each new cohort
(i.e., patient 4, 7, 10, 13, 16, \ldots) was enrolled.

The three patients in cohort 1 were enrolled at days 3, 6 and 18, respectively
(see Table S6 in the supplementary materials [\citet{supp}]). On day 28 when patient 4
(i.e., the first patient of cohort 2) was enrolled, the three patients in
cohort 1 had finished only 28\%, 24\% and 11\% of their 3-month follow-ups
without experiencing toxicity (i.e., DLT). The estimates of
the toxicity probabilities of six dose levels were
$\hat{\pi} = (0.172, 0.185, 0.209, 0.236, 0.264, 0.294)$.
We escalated the dose and treated patient 4,
and subsequently patients 5 and 6, at dose level 2.

When patient 7 (the first patient of cohort 3) arrived on day 57, we
again updated the estimates of the
toxicity probabilities and obtained $\hat{\pi} = (0.315, 0.336, 0.369,
0.407, 0.445, 0.486)$.
Although at that moment we still had not observed any DLTs yet,
the values of $\hat{\pi}$ increased compared with
the previous estimates of $\pi$. This is because on day 57,
more patients (i.e., patients 1 to 6) were under treatment and none of
them had
finished their 3-month follow-ups yet. There was greater uncertainty
regarding the toxicity probabilities of the doses and it was preferable to
be conservative. Our algorithm automatically took into account such
uncertainty and
de-escalated the dose back to the first level for treating cohort 3.

On day 91 when the first patient of cohort 4 (i.e., patient 10) was accrued,
patients 1, 2 and 3 were very close to completing
their 3-month follow-ups without experiencing toxicity,
indicating that the first dose level was safe and dose escalation was needed.
The proposed algorithm timely reflected this data information and
escalated the dose to level 2. The dose was further escalated to levels~3 and~4 for treating cohorts 5 and 6, respectively, as no DLT was observed.
By the time patient 19 arrived, the toxicity outcomes of all patients
treated in the trial had been observed. In particular, all three
patients (i.e., patients 16--18) treated at dose level~4 had
experienced DLTs.
Our algorithm de-escalated the dose to level 3 to treat cohort 7.
Thereafter, there were always at least 18 toxicity outcomes (from
patients 1--18) fully observed, thus, $\hat{\pi}$
became rather stable and consistently indicated that dose
3 was the MTD. The last 3 cohorts
were all treated at dose level 3 and, at the end of the trial, dose
3 was selected as the MTD with the estimated toxicity probability of 0.259.

\begin{figure}

\includegraphics{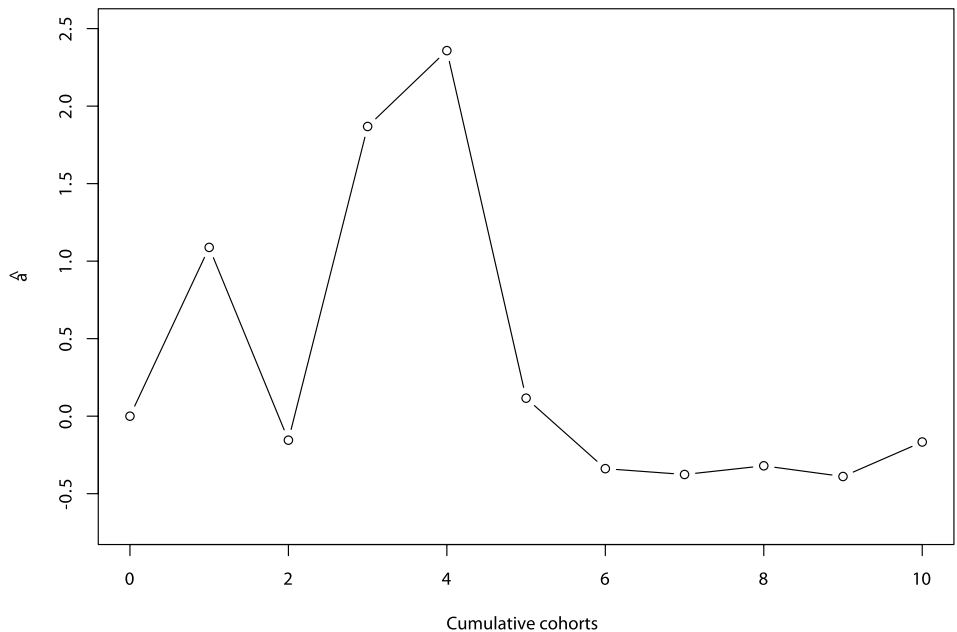}

\caption{Estimate of the unknown parameter $a$
with cumulative cohorts in the esophageal cancer trial.} \label{app1fig}
\end{figure}

Figure~\ref{app1fig} displays the estimate of
the unknown parameter $a$ during the trial conduct.
At the beginning of the trial, there was much variability
for the estimate of $a$ due to sparse data, while
the estimate became stabilized after six cohorts of
patients were enrolled. Correspondingly,
Table~\ref{app1est} summarizes the estimates of the
toxicity probabilities $\pi$ for the six doses at each
decision-making time.

\begin{table}
\tabcolsep=0pt
\caption{Estimates of the toxicity probabilities of six doses with
the cumulative number of cohorts in the esophageal cancer trial}\label{app1est}
\begin{tabular*}{\textwidth}{@{\extracolsep{\fill}}lcccccccccc@{}}
\hline
&\multicolumn{10}{c@{}}{\textbf{Cumulative number of cohorts}}\\[-6pt]
&\multicolumn{10}{c@{}}{\hrulefill}\\
\textbf{Dose level} & \textbf{1} & \textbf{2} & \textbf{3} & \textbf{4}
& \textbf{5} & \textbf{6} & \textbf{7} & \textbf{8} & \textbf{9} &
\textbf{10} \\
\hline
1 & 0.172 & 0.315 & 0.028 & 0.021 & 0.080 & 0.175 & 0.186 & 0.171 &
0.189 & 0.125 \\
2 & 0.185 & 0.336 & 0.035 & 0.026 & 0.114 & 0.228 & 0.240 & 0.223 &
0.243 & 0.172 \\
3 & 0.209 & 0.369 & 0.050 & 0.037 & 0.181 & 0.320 & 0.333 & 0.314 &
0.337 & 0.259 \\
4 & 0.236 & 0.407 & 0.071 & 0.053 & 0.268 & 0.422 & 0.435 & 0.415 &
0.440 & 0.361 \\
5 & 0.264 & 0.445 & 0.096 & 0.071 & 0.359 & 0.515 & 0.528 & 0.509 &
0.532 & 0.458 \\
6 & 0.294 & 0.486 & 0.128 & 0.093 & 0.454 & 0.603 & 0.615 & 0.598 &
0.619 & 0.553 \\
\hline
\end{tabular*}
\end{table}

\section{Conclusions}\label{sec5}
We have proposed the DA-CRM design to address the issues associated with
late-onset toxicities in phase I dose-finding trials.
In the new design, unobserved toxicity outcomes are
naturally treated as missing data. We established that
such missing data are nonignorable and linked the missing
data mechanism with the time to toxicity based on a flexible piecewise
exponential model.
Simulation studies showed that the DA-CRM outperforms other
available methods, particularly when toxicities need a long follow-up time
to be assessed. The selection percentage of the DA-CRM is
often close to the optimal value, and many
fewer patients would be treated at overly toxic doses.

This paper has focused on the single-agent dose finding using the CRM,
but the proposed methodology provides a general and systematic approach
to transforming the late-onset toxicity problem into a standard
complete-data problem by imputing the missing toxicity outcomes. The
proposed method can serve as a universal adaptor to extend existing
trial designs to accommodate more complicated dose-finding problems
with late-onset toxicity. For example, by incorporating the data
augmentation procedure into the partial-order CRM [Wages, Conaway and
O'Quigley (\citeyear{WagConOQu11})], we can address the late-onset toxicity for
drug-combination trials or dose finding with group heterogeneity.
It is also worth emphasizing that although we
have focused on the late-onset toxicity, the proposed method can also be
used to handle other kinds of late-onset outcomes, such as
delayed efficacy responses in phase I/II or phase II trials, as well as
response-adaptive randomization designs.


\begin{appendix}\label{app}

\section*{\texorpdfstring{Appendix: Proof of Theorem \lowercase{\protect\ref{th1}}}
{Appendix: Proof of Theorem 1}}

Considering that each subject is fully followed up to
$T$, if $t_i > T$, then $Y_i=0$; and if $t_i \le T$, then $Y_i=1$.
We demonstrate the nonignorable missingness for the
missing data caused by late-onset toxicity as follows. For a subject
who will
not experience toxicity, the probability that his/her toxicity outcome
will be
missing is given by
\begin{eqnarray*}
\Pr(M_i=1|Y_i=0) &=& \Pr(t_i>u_i,
u_i<T|Y_i=0)
\\
&=& \Pr(u_i<T|Y_i=0)\Pr(t_i>u_i
|u_i<T, Y_i=0)
\\
&=& \Pr(u_i<T|t_i>T)\Pr(t_i>u_i
|u_i<T, t_i>T)
\\
&=& \Pr(u_i<T),
\end{eqnarray*}
where the last equality
follows because $t_i$ and $u_i$ are independent, and $\Pr(t_i>u_i
|u_i<T, t_i>T)=1$. Similarly,
for a subject who will experience toxicity, the probability that
his/her toxicity outcome will be
missing is given by
\begin{eqnarray*}
\Pr(M_i=1|Y_i=1) &=& \Pr(t_i>u_i,
u_i<T|Y_i=1)
\\
&=& \Pr(u_i<T|Y_i=1)\Pr(t_i>u_i
|u_i<T, Y_i=1)
\\
&=& \Pr(u_i<T|t_i\le T)\Pr(t_i>u_i
|u_i<T, t_i\le T)
\\
&=& \Pr(u_i<T)\Pr(t_i>u_i
|u_i<T, t_i \le T).
\end{eqnarray*}
Because of $\Pr(t_i>u_i |u_i<T, t_i\le T)<1$, it follows that
\[
\Pr(M_i=1|Y_i=0) > \Pr(M_i=1|Y_i=1).
\]
Therefore, the missing data are more likely to occur for
those patients who will not experience toxicity in $(0, T)$.
\end{appendix}

\section*{Acknowledgments}
The authors thank the Editor Professor Kafadar, Associate Editor and
three anonymous
referees for their many constructive and insightful comments that have
resulted in significant
improvements in the article.

\begin{supplement}[id=suppA]
\stitle{Additional simulation results}
\slink[doi]{10.1214/13-AOAS661SUPP} 
\sdatatype{.pdf}
\sfilename{aoas661\_supp.pdf}
\sdescription{Additional simulation results.}
\end{supplement}

%

\printaddresses

\end{document}